\documentclass[letterpaper,twocolumn,english,pra,aps,amsmath,amssymb,superscriptaddress,floatfix]{revtex4}
\usepackage[T1]{fontenc}
\usepackage[latin9]{inputenc}
\usepackage{color}
\usepackage{amsmath}
\usepackage{amssymb}
\usepackage{graphicx}
\usepackage{esint}

\makeatletter

\@ifundefined{textcolor}{}
{%
 \definecolor{BLACK}{gray}{0}
 \definecolor{WHITE}{gray}{1}
 \definecolor{RED}{rgb}{1,0,0}
 \definecolor{GREEN}{rgb}{0,1,0}
 \definecolor{BLUE}{rgb}{0,0,1}
 \definecolor{CYAN}{cmyk}{1,0,0,0}
 \definecolor{MAGENTA}{cmyk}{0,1,0,0}
 \definecolor{YELLOW}{cmyk}{0,0,1,0}
 }

\usepackage[dvips]{epsfig}
\usepackage{float}\usepackage{amsfonts}
\usepackage{euscript}
\usepackage{enumerate}\usepackage{hhline}\usepackage{threeparttable}
\usepackage{pslatex}\usepackage{tabularx}\@ifundefined{definecolor}
 {\usepackage{color}}{}

\usepackage{dcolumn}
\usepackage{bm}

\renewcommand{\@biblabel}[1]{#1. }
\renewcommand{\@dotsep}{500}
\renewcommand{\@pnumwidth}{0em}
\renewcommand{\l@figure}[2]{
        \@dottedtocline{1}{1.5em}{2em}{Figure #1}{}\vspace{15pt}}

\newcommand{\comment}[1]{}

\makeatother

\usepackage{babel}
\begin{document}

\title{Quantum light generation on a silicon chip using waveguides and resonators}
\author{Jun Rong Ong}
\email{j5ong@ucsd.edu}
\author{Shayan Mookherjea}
\email{smookherjea@ucsd.edu}
\affiliation{University of California, San Diego, Mail Code 0407, La Jolla, California 92093}
\begin{abstract} Integrated optical devices may replace bulk crystal or fiber based assemblies with a more compact and controllable photon pair and heralded single photon source and generate quantum light at telecommunications wavelengths. Here, we propose that a periodic waveguide consisting of a sequence of optical resonators may outperform conventional waveguides or single resonators and generate more than 1 Giga-pairs per second from a sub-millimeter-long room-temperature silicon device, pumped with only about 10 milliwatts of optical power. Furthermore, the spectral properties of such devices provide novel opportunities of wavelength-division multiplexed chip-scale quantum light sources.  
\end{abstract}
\date{\today}

\maketitle
\begin{figure}
\textcolor{black}{\includegraphics[width=1\columnwidth]{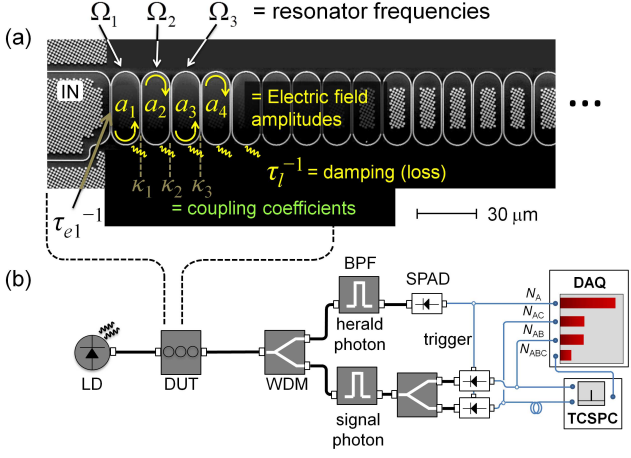}
\caption{(a) A coupled resonator waveguide consisting of $N$ directly-coupled microring resonators. The waveguide eigenmode is a Bloch excitation, i.e., a collective oscillation of all $N$ resonators, with a fixed relationship between adjacent resonators \cite{Notomi2008}. The direction of light circulation in each resonator is as indicated for the specified input. In the notation used in this paper, the field operator of successive resonators are $a_{1}$, $a_{2}$, $a_{3}$, ..., the resonance radial frequencies are $\Omega_{1}$, $\Omega_{2}$, $\Omega_{3}$, ..., the inter-ring coupling coefficients are labeled $\kappa_{2}$, $\kappa_{3}$, $\kappa_{4}$, ..., and the input/output external coupling coefficients are labeled by their rates $\frac{1}{\tau_{e1}}$ and $\frac{1}{\tau_{e2}}$ (the latter is not shown, at the output side of the chip). The resonator loss is indicated by the damping rate $\frac{1}{\tau_{l}}$. (b) Light from a laser diode (LD) is coupled into the waveguide device under test (DUT), and output photon pairs are spectrally separated (wavelength division multiplexer, WDM), filtered (band pass filter, BPF), and detected e.g., by single photon avalanche detectors (SPADs) shown here in the configuration needed to perform a $g^{(2)}(0)$ measurement using a time-correlated single photon counting system (TCSPC) and a data acquisition card (DAQ). \label{fig:1}}}
\end{figure}

Trends in quantum optics are evolving towards chip-scale photonics \cite{OBrien2009}, with one of the eventual goals being the full-fledged
combination of sources, circuits, and detectors on a single chip. Regarding chip-scale sources, researchers have predicted and shown that an optically-pumped spontaneous four-wave mixing (SFWM) process in silicon can be used to generate entangled photon pairs in waveguides and resonators \cite{LinAgrawal2006,Sharping2006,Harada2008,Clemmen2009}. This third-order nonlinear process is similar to the second-order spontaneous nonlinear optical processes induced in bulk optical crystals (except scaling with the square of the pump power instead of linearly), and before being investigated in lithographically-fabricated waveguides, has been demonstrated in optical fiber \cite{Grangier1986,Fiorentino2002}. As a further step, we have explicitly shown heralded single photons at 1.55 $\mathrm{\mu m}$ wavelength from a silicon chip at room temperature (see Fig. \ref{fig:1}) \cite{davanco2012}. Given the maturity of integrated optics technology, it is realistic to envision on-chip high-brightness single-photon sources at wavelengths compatible with the worldwide fiber optic internet infrastructure.  

However, an important open question is: What is the optimal device for generating quantum light using an integrated photonic structure? To be specific, we focus on devices made using silicon. The photon pair generation rate depends on the intrinsic four-wave mixing nonlinear coefficient ($\gamma=2\pi n_{2}/\lambda A_{\text{eff}}$),
in terms of the Kerr nonlinear index $n_{2}$ and the effective area of the waveguide mode ($A_{\text{eff}}$), the waveguide length ($L$),
the pump power ($P$), and the loss coefficient of lithographically-fabricated
waveguides ($\alpha$). Silicon nanophotonic waveguides are already quite promising, compared to optical fiber or bulk crystals, since a single mode `` wire'' waveguide with cross-sectional dimensions of about 0.5 x 0.25 $\mathrm{\mu m}^{2}$
has a nonlinearity coefficient $\gamma_{\text{Si}}$= 100-200 $\text{W}^{-1}\text{m}^{-1}$ (five orders of magnitude greater than optical fiber)
around a wavelength of 1.5 $\mathrm{\mu m}$ \cite{Osgood2009}. But chip-scale devices present special challenges as $L$ is limited to only a few
centimeters within a typical die site, and on-chip footprint is a highly valuable resource in CMOS fabrication. Moreover, for a waveguide
that is fabricated with loss coefficent $\alpha$, the effective interaction length of nonlinear interactions $L_{\text{eff}}=[1-\exp(-\alpha L)]/\alpha$
can be significantly smaller than $L$ when $\alpha L\ge1$. Also, pump powers $P$ in silicon are limited to a few milliwatts to minimize the probability of multi-photon generation and avoid two-photon absorption and free-carrier generation losses. 

The indistinguishability of output single photons is also an important consideration \cite{Fulconis2007}. In silicon
waveguides, the phase-matching bandwidth of the SFWM process is generally quite broad, on the order of tens of nanometers. As such, the generated photon pair emerges as an entangled state, and detection of the heralding photon projects the signal
photon into a mixed state. Purity may be enhanced by spectrally filtering the output, the disadvantage being a reduction in photon count rate since unused pairs are discarded. Recent work \cite{Garay-Palmett:07} has also
shown that through the careful control of waveguide dispersion, photon
pairs may be generated in factorable states which are spectrally de-correlated.
Alternatively, one may limit the modes available for SFWM process
by placing the nonlinear material in a cavity, thereby providing both
spectral filtering of output states as well as local intensity enhancement
of the pump\cite{Lu2000,Moreno2010}. 

Based on these considerations, we study whether
a micro-resonator is preferable to a conventional waveguide as a heralded
single photon source with specific reference to silicon devices. We
then show that a particular type of hybrid device {[}Fig. \ref{fig:1}{]},
which consists of a linear array of nearest-neighbor coupled microresonators,
can possibly generate in excess of 1 Giga-pairs per second
for 10-20 mW of optical pump power, from a waveguide that is only 0.1~mm long,
thus outperforming existing photon pair sources by 1-2 orders of magnitude
in generation rates and by 2-3 orders of magnitude in device size.

In single ring resonators, the theory of both parametric downconversion (second order nonlinearity) as well as SFWM (third order nonlinearity) has been studied \cite{Scholz2009,Helt2010,Chen2011}. Here we extend the previously described methods to develop the output
state of the photon pair from a series of directly coupled rings, so that waveguides, rings and coupled-ring waveguides can be compared. We begin with the phenomenological Hamiltonian,
\begin{equation}
\begin{array}{c}
H=\underset{m}{\sum}\underset{l=s,i}{\sum}\hbar\Omega_{m}a_{l,m}^{\dagger}a_{l,m}+\hbar\kappa_{l,m}a_{l,m}^{\dagger}a_{l,m-1}\\
+\hbar\kappa_{l,m+1}a_{l,m}^{\dagger}a_{l,m+1}+\hbar\chi_{m}a_{s,m}^{\dagger}a_{i,m}^{\dagger}
\end{array}
\end{equation}
where $a_{l,m}^{\dagger}$ are the field operators of the resonator
modes $l=s,i$ at the resonator site $m$, $\Omega_{m}$ are the resonance
frequencies, $\kappa_{l,m+1}$ are the inter-resonator coupling coefficients
and $\chi_{m}$ is the coefficient proportional to the Kerr nonlinearity.
In general $\chi_{m}$ may be time-dependent, $\chi_{m}(t)=\frac{\gamma_{0}v_{g}}{T_{c}}[A_{p,m}(t)]^{2}$,
where $A_{p,m}(t)=a_{p,m}(t)e^{-i\Omega_{p}t}$ is the classical pump
with a slowly-varying amplitude at carrier-frequency $\Omega_{p}$,
$\gamma_{0}$ is the usual waveguide nonlinear parameter, $v_{g}$
is the waveguide group velocity, $T_{c}=1/FSR$ is the round-trip
time (inverse of the free-spectral range). We adopt the approach of
Collett and Gardiner\cite{Gardiner1985} (i.e., time-domain coupled mode theory) to obtain the equations of motion in the Heisenberg
picture. In the single resonator case, these may be written explicitly as, \begin{subequations}
\begin{equation}
[\frac{1}{\tau_{s}}-i(\omega_{s}-\Omega_{s})]a_{s}(\omega_{s})=-i\int\chi(\omega_{s}+\omega_{i})a_{i}^{\dagger}(\omega_{i})d\omega_{i}-i\mu a_{s,in}
\end{equation}
\begin{equation}
[\frac{1}{\tau_{i}}+i(\omega_{i}-\Omega_{i})]a_{i}^{\dagger}(\omega_{i})=+i\int\chi^{\dagger}(\omega_{s}+\omega_{i})a_{s}(\omega_{s})d\omega_{s}+i\mu a_{i,in}^{\dagger}
\end{equation}
\end{subequations}where $a_{s}(\omega_{s})$ are the frequency components
of the time-dependent field operator $a{}_{s}(t)$ and $\frac{1}{\tau_{s}}=\frac{1}{\tau_{l}}+\frac{1}{\tau_{e}}$
is the damping coefficient which includes effects of loss and external
coupling. These equations contain the same information as the joint-spectral
amplitude, modified by the cavity enhancement effects. In the quasi-cw
limit, one may forgo the integral and solve the coupled equations
as was done in \cite{Chuu2011}.\begin{subequations} 
\begin{equation}
a_{out,s}(\omega_{s})=-\mu^{2}[A(\omega_{s},\omega_{i})a_{in,s}(\omega_{s})+B(\omega_{s},\omega_{i})a_{in,i}^{\dagger}(\omega_{i})]
\end{equation}
\begin{equation}
a_{out,i}^{\dagger}(\omega_{i})=-\mu^{2}[C(\omega_{s},\omega_{i})a_{in,s}(\omega_{s})+D(\omega_{s},\omega_{i})a_{in,i}^{\dagger}(\omega_{i})]
\end{equation}
\end{subequations}We have used the boundary condition $|a_{out}|^{2}=\mu^{2}|a|^{2}$,
where $\mu^{2}=\frac{2}{\tau_{e}}$ is the input mode coupling coefficient.
In the case of vacuum input and low gain the power spectral density
of the output photons, $\sigma(\omega_{s},\omega_{i})=<a_{out,s}^{\dagger}a_{out,s}>=\frac{\mu^{4}|\chi(\omega_{s}+\omega_{i})|^{2}}{|\frac{1}{\tau_{s}}-i(\omega_{s}-\Omega_{s})|^{2}|\frac{1}{\tau_{i}}+i(\omega_{i}-\Omega_{i})|^{2}}$
and the total signal flux is $F=\frac{1}{2\pi}\int\sigma(\omega)d\omega$,
where the idler frequency is implicitly related by the energy conservation
$2\omega_{p}=\omega_{s}+\omega_{i}$. Alternatively, by taking $\chi(\omega_{s}+\omega_{i})$
as the pump distribution in the pulsed pump regime, $\sigma(\omega_{s},\omega_{i})$
is interpreted as the joint spectral intensity. 

Extending to the case of multiple coupled cavities\cite{Poon:07},
we have the following matrix equation,
\begin{subequations}
\begin{equation}
\left[\begin{array}{c}
a_{s,1}\\
a_{s,2}\\
\vdots\\
a_{i,1}^{\dagger}\\
a_{i,2}^{\dagger}\\
\vdots
\end{array}\right]_{2N\times1}=-i\mu\vec{T}\left[\begin{array}{c}
a_{s,in}\\
0\\
\vdots\\
a_{i,in}^{\dagger}\\
0\\
\vdots
\end{array}\right]_{2N\times1}\label{eq:4}
\end{equation}

\begin{equation}
\vec{T}=\left[\begin{array}{cc}
M_{s} & C\\
C^{\dagger} & M_{i}
\end{array}\right]_{2N\times2N}^{-1}
\end{equation}
\begin{widetext}
\begin{equation}
M_{s}=\left[\begin{array}{ccccc}
-i(\omega_{s}-\Omega_{s,1})+\frac{1}{\tau_{l}}+\frac{1}{\tau_{e1}} & -i\kappa_{s,2} & 0 & \cdots & 0\\
-i\kappa_{s,2} & -i(\omega_{s}-\Omega_{s,2})+\frac{1}{\tau_{l}} & . & \cdots & 0\\
0 & . & . & . & .\\
\vdots & \vdots & 0 & . & -i(\omega_{s}-\Omega_{s,N})+\frac{1}{\tau_{l}}+\frac{1}{\tau_{e2}}
\end{array}\right]_{N\times N}
\end{equation}
\end{widetext}
\begin{equation}
C=\left[\begin{array}{cccc}
-i\chi_{1} & 0 & \cdots & 0\\
0 & -i\chi_{2} & \cdots & 0\\
0 & 0 & \ddots & .\\
\vdots & \vdots & . & -i\chi_{N}
\end{array}\right]_{N\times N}
\end{equation}
\end{subequations}
and we have assumed a single sided input/output.

Similar to single ring case, we have for the coupled-resonator waveguide, 
\begin{subequations}
\begin{multline}
a_{out,s}(\omega_{s})=-\mu_{1}\mu_{2}[T_{N,1}(\omega_{s},\omega_{i})a_{in,s}(\omega_{s}) \\ 
+T_{N,N+1}(\omega_{s},\omega_{i})a_{in,i}^{\dagger}(\omega_{i})]
\end{multline}
\begin{multline}
a_{out,i}^{\dagger}(\omega_{i})=-\mu_{1}\mu_{2}[T_{2N,1}(\omega_{s},\omega_{i})a_{in,s}(\omega_{s})\\ 
+T_{2N,N+1}(\omega_{s},\omega_{i})a_{in,i}^{\dagger}(\omega_{i})]
\end{multline}
\end{subequations} and the joint spectral intensity $\sigma(\omega_{s},\omega_{i})=\mu_{1}^{2}\mu_{2}^{2}|T_{N,N+1}|^{2}$.
We note here that the coupled mode theory result is equivalent to
the first-order perturbation theory with a cavity modified joint spectral
amplitude, $\vert\psi\rangle={\vert0\rangle}_{s}{\vert0\rangle}_{i}+g\iint d\omega_{s}d\omega_{i}\,\textrm{FE}_{s}(\omega_{s})\textrm{FE}_{i}(\omega_{i})\textrm{FE}_{p}(\omega_{s},\omega_{i}) \times f(\omega_{s},\omega_{i})a^{\dagger}(\omega_{s})a^{\dagger}(\omega_{i}){\vert0\rangle}_{s}{\vert0\rangle}_{i}$
where the subscripts $s$ and $i$ refer to the signal and idler frequencies,
$g$ is proportional to the photon-pair production rate, and the function
$f(\omega_{s},\omega_{i})$ which describes the phase-matching and
pump spectral envelope, is the joint spectral amplitude \cite{Moreno2010}.
FE are the field enhancement factors and are equivalent to the slowing
factors used in \cite{Ong2011}. 
\begin{figure}
\includegraphics[width=0.95\columnwidth]{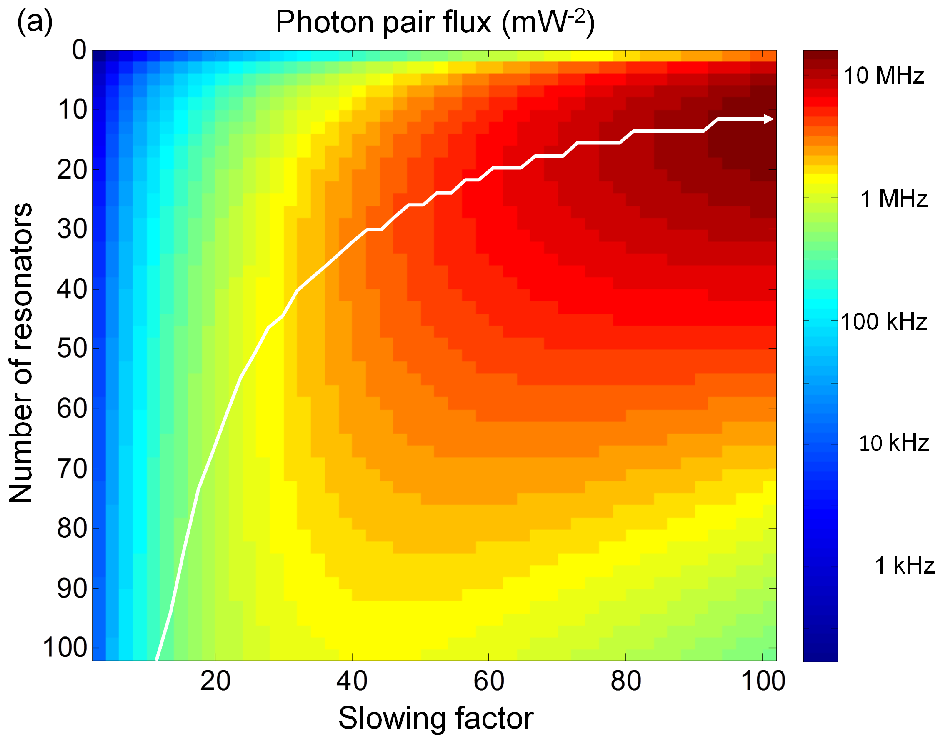}

\includegraphics[width=0.95\columnwidth]{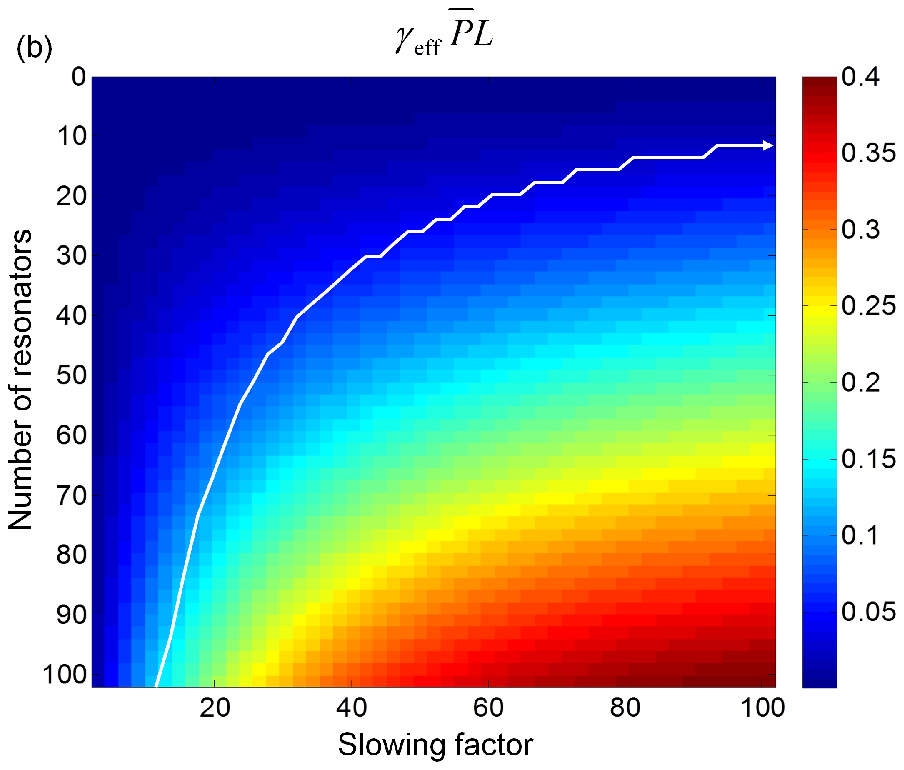}

\includegraphics[width=0.95\columnwidth]{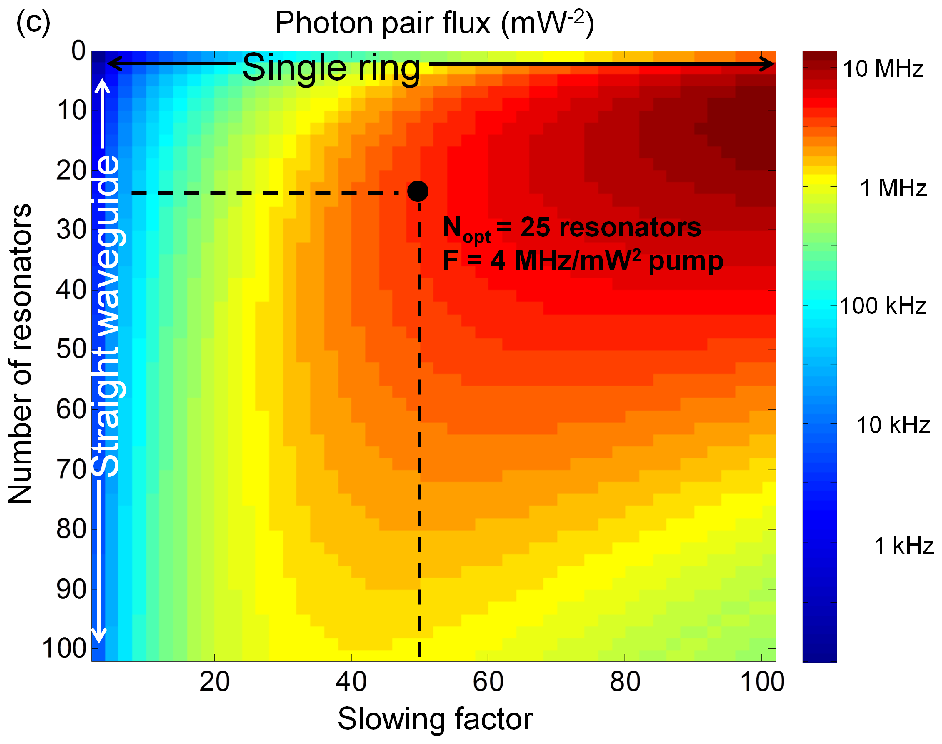}\caption{(a) Calculated photon pair flux $\mathit{F}$ using pair generation
equations, Eq. (\ref{eq:6}). The white trend-line follows the optimum
number of resonators for a given slowing factor. (b) Corresponding
values of $\gamma_{\textrm{eff}}\overline{P}L$ for each $\mathit{S}$
and $\mathit{N}$, showing the low multiphoton generation probability along the white line. (c) Calculated photon pair flux $\mathit{F}$ using
coupled mode equations, Eq. (\ref{eq:4}). The top region of the contour
plot represents a single resonator, while the far left approaches
that of a conventional silicon nanowire waveguide. For $S=50$, the
optimum number of resonators is $N_{opt}=25$ for which $F=4\textrm{ MHz}/\textrm{mW}^{2}$.
\label{fig:2}}

\end{figure}

We verify the agreement between the time-domain coupled mode equations
and the slowing factor enhanced pair generation equations by comparing
the calculated pair flux. In the discussion below, we will assume
a simplified picture with flat spectral filtering about the desired
signal and idler modes, as was done in previous experiments \cite{Fulconis2005}.
The number of photon pairs generated per second is given in the low
pump power regime by 
\begin{equation}
F=\Delta\nu\left(\gamma_{\textrm{eff}}PL_{\text{eff}}\right)^{2}\,\exp(-\alpha L)\,\label{eq:6}
\end{equation}
 where $\gamma_{\textrm{eff}}^{2}=S_{s}S_{i}{\left(\frac{S_{p}+1}{2}\right)}^{2}\gamma_{0}^{2}$,
$S_{\{p,s,i\}}$ are the slowing factors at the pump ($p$), signal
($s$) and idler ($i$) wavelengths, and $L_{\text{eff}}=[1-\exp(-\alpha L)]/\alpha$
represents an effective propagation length, defined as the geometric
length $L=N\pi R$ renormalized by the loss coefficient, $\alpha$.
An experimentally-validated transfer-matrix method can be used to
calculate the $\alpha$ coefficient which scales linearly with the
slowing factor \cite{CooperPTL2011}. We assume that the linear loss coefficienct $\alpha$ does not vary significantly with wavelength over the bandwidth of interest. To account for nonlinear absorption losses in silicon \cite{Osgood2009} we substitute $\alpha\rightarrow\alpha+2\frac{\overline{P_{p}}}{A_{\textrm{eff}}}\beta L$
and $PL_{\textrm{eff}}\rightarrow\overline{P}L$ where $\overline{P}=[\log(1+\frac{\beta}{A\textrm{eff}}PL_{\textrm{eff}})]/\frac{\beta}{A\textrm{eff}}L$
and $\beta$ is the effective TPA coefficient of the coupled-resonator waveguide which
scales in the same way as $\gamma_{\textrm{eff}}$ with $S$, i.e.
$\beta\varpropto S^{2}\beta_{0}$. For an apodized structure, which we
define as the case where the boundary coupling coefficients are matched
to the input/output waveguides \cite{Liu:11}, we have at resonance
$S=1/|\kappa|$, where $|\kappa|$ is the inter-resonator coupling
coefficient in the transfer-matrix formalism. The bandwidth of the
photon generation process, $\Delta\nu$, is assumed to be the linewidth
of a Bloch eigenmode of the coupled-resonator waveguide, which scales inversely with the
number of resonators in the chain, $N$, 
\begin{equation}
\Delta\nu\approx\frac{1}{N}\frac{2\textrm{FSR}}{\pi}\sin^{-1}|\kappa|.\label{eq:7}
\end{equation}

Calculations were performed using the following parameters, $R=5\ \mathrm{\mu m}$,
waveguide loss = 1 dB/cm, $\gamma_{0}=200\ \text{W}^{-1}\text{m}^{-1}$,
$\beta_{0}=0.75\ \textrm{cm/}\textrm{GW}$, $P=1\ \textrm{mW}$ to
obtain $F$ over a range of values of $S$ and $N$, showing good
agreement between the pair generation equations and coupled mode equations
{[}Fig. \ref{fig:2} (a) and (c) respectively{]}. Resonator chains
that are in excess of the optimum length, or with too high a value
of $S$ incur penalties because of the exponential loss factor in
Eq. (\ref{eq:6}), and the collapse of the bandwidth $\Delta\nu$.
Too small values of $S$ do not fully utilize the slow-light enhancement
of the nonlinear FWM coefficient, which scales as a higher
power of $S$ than the corresponding decrease of bandwidth, unlike
in a (linear) slow-light delay line. The optimum parameters are large
$S$ and small $N$, i.e. towards the single resonator configuration,
for which the maximum pair flux rate exceeds 10 MHz at 1 mW pump power (and scaling quadratically with the pump power). 

\textit{Scaling difference between single rings and coupled ring waveguides:} One of the questions regarding the optimum device geometry for generating photon pairs is the appropriate size of resonators. Recently, the efficiency of classical and spontaneous four-wave mixing in single microring resonators has been compared \cite{Helt:2012,Azzini2012}, with the conclusion being that in both cases, the conversion efficiency scales with the ring radius as $R^{-2}$, i.e., smaller rings are better than larger rings in generating photon pairs. This results from the analytically derived expression for the spontaneously-generated idler power $P_{i,SP}$ (from an injected pump power $P_{p}$ at optical carrier frequency $\omega_{p}$) 
\begin{equation}
P_{i,SP}=(\gamma 2 \pi R)^{2} \left( \frac{Qv_{g}}{\omega_{p} \pi R} \right)^{3} \frac{\hbar \omega_{p} v_{g}}{4 \pi R}P_{p}^{2},\label{eq:8}
\end{equation}
and a key assumption, that the ring quality factor $Q$ is independent of the ring radius $R$. Starting with the equation for the (loaded) quality factor of a ring resonator side-coupled to a waveguide \cite{Steier2009}, 
\begin{subequations}\label{eq:9}
\begin{equation}
Q = \frac{\pi\sqrt{a_{rt}\tau}}{1-a_{rt}\tau}\frac{n_{g}L}{\lambda}
\end{equation}
where $a_{rt}= \textrm{exp}(-\alpha L/2)$ and $L = 2\pi R$, we examine two limiting cases as examples. In the first case, we examine a weakly coupled resonator ($\tau = \sqrt{1-|\kappa|^{2}} = 1$) with low loss ($a_{rt} \approx 1 -\alpha L/2$) in which case the quality factor can be expressed as, 
\begin{equation}
Q = \frac{2\pi n_{g}}{\lambda\alpha}
\end{equation}
which is the intrinsic $Q$ limit. In this case, $Q$ is indeed independent of $R$, and $P_{i,SP}$ scales as  $R^{-2}$. In the second case, however, we assume that the loaded $Q$ is dominated by the coupling coefficient ($|\kappa| \neq 0$) and then
\begin{equation}
Q = \frac{2\pi n_{g}}{\lambda|\kappa|^{2}/L}.
\end{equation} 
\end{subequations}
In this case, the ratio $Q/R$ in Eq. (\ref{eq:8}) is length-invariant, and $P_{i,SP}$ increases linearly with $R$. As previously shown \cite{MookherjeaSchneiderOL2011}, coupled-resonator waveguides are more disorder tolerant in the large-coupling regime, and therefore, Eq. (\ref{eq:9}c) is more appropriate in describing performance rather than Eq. (\ref{eq:9}b). In fact, the agreement between Fig. \ref{fig:2}(a), calculated using the conventional waveguide theory with nonlinearities scaled by the slowing factor, and Fig. \ref{fig:2}(c), calculated using the first-principles time-domain coupled-mode theory model, shows that coupled-resonator waveguides are, in fact, more similar to waveguides than single resonators in many ways, with the attendant benefits of a slowing factor in enhancing the nonlinearity per unit length. Here, it is useful to recall, as shown in the classical domain, that coupled-resonator waveguides break the traditional trade-offs between parametric conversion efficiency and bandwidth, and are more robust against chromatic dispersion and propagation loss, compared to conventional waveguides \cite{Morichetti-NC-2011}. Similarly, in the quantum domain, coupled-resonator waveguides may outperform conventional waveguides as pair and heralded single photon sources. 

\textit{Multi-photon generation probability:} For a heralded single photon source we require low multi-photon probability. The inset
of Fig. \ref{fig:2}(a) shows the value of the quantity $\gamma_{\textrm{eff}}\overline{P}L$
for each value of $S$ and $N$. For a $\gamma_{\textrm{eff}}\overline{P}L\ll1$,
the level of stimulated scattering events is kept relatively low \cite{LinAgrawal2006}
which is true for the regions of highest pair flux (large $S$ and
small $N$). 

\begin{figure}
\includegraphics[width=1\columnwidth]{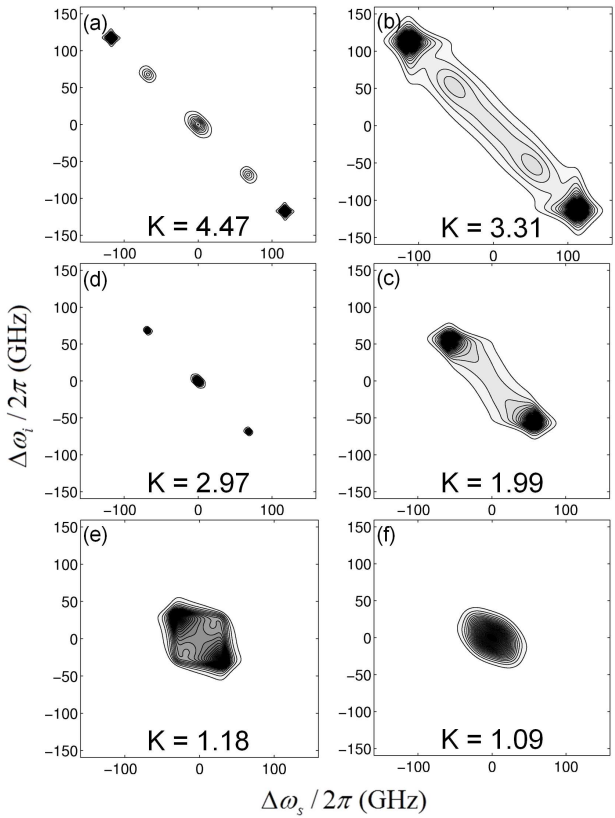}
\caption{Joint Spectral Intensity (JSI) plots for various coupling coefficient
configurations, assuming that the coupling coefficients between adjacent resonators, shown in Fig. 1, can be individually altered. (a) Unapodized (b) Apodized (c),(d) Chosen from a sample of Monte Carlo simulations with random coupling coefficients. (e) JSI for coupling coefficients chosen so as to realize a Butterworth filter response and (f) Bessel filter response in the linear transmission regime.\label{fig:3}}
\end{figure}

\textit{Joint Spectral Intensity (JSI):} To evaluate the spectral characteristics of the signal-idler photon pair, we calculate the Joint Spectral Intensity, and also the Schimdt number $K=1/\sum\lambda^{2}$, which is the sum of
the squares of the Schmidt eigenvalues (for a pure state $K=1$)
\cite{Law2004}. In Fig. \ref{fig:3}(a), (b), we plot the joint spectral
intensities of an unapodized and apodized coupled-resonator waveguide of similar inter-resonator
coupling coefficients. The shape of the spectrum reflects the number
of resonators chosen $N=5$, with the peaks corresponding to the locations
of maximum transmission, which are also the Bloch eigenmodes. The
pump pulse width is taken as 10 ps in both cases and we obtain $K=4.47$
for the unapodized device and $K=3.31$ for the apodized device. However,
we note that choosing shorter pulses does not significantly change
the Schmidt number in contrast with the single ring case \cite{Helt2010}.
In order to herald pure state single photons, filtering will be necessary.
Choosing a filter bandwidth equal to the Bloch eigenmode width given
by Eq. (\ref{eq:7}), we are able to obtain approximately a single
Schimdt mode output. 

On the other hand, if we have control over each individual inter-resonator
coupling coefficients we are able to synthesize a large variety of
different joint spectral amplitudes with different Schimdt numbers.
In Fig. \ref{fig:3}(c), (d) we plot two interesting contours taken
from a sample of different inter-resonator coupling configurations,
each coefficient being a pseudo-random number ranging from 0 to 1.
Clearly, with the added control over individual couplers we can obtain
a large variety of corresponding $K$ values. Of special interest
are the configurations giving maximally flat transmission (Butterworth)
and maximally flat group delay (Bessel) \cite{Liu:11} since these
quantities define the overall shape of the output joint spectrum {[}see
Fig. \ref{fig:3}(e), (f){]}. Without additional filtering, we are
able to obtain close to a pure heralded state for both the Butterworth
filter configuration ($K=1.18$) and the Bessel filter configuration
($K=1.09$). Of course, filtering will still be required before the detectors, to separate the signal and idler photons and reject any unused pumps from reaching the SPADs \cite{davanco2012}. 

\begin{figure}
\includegraphics[width=1\columnwidth]{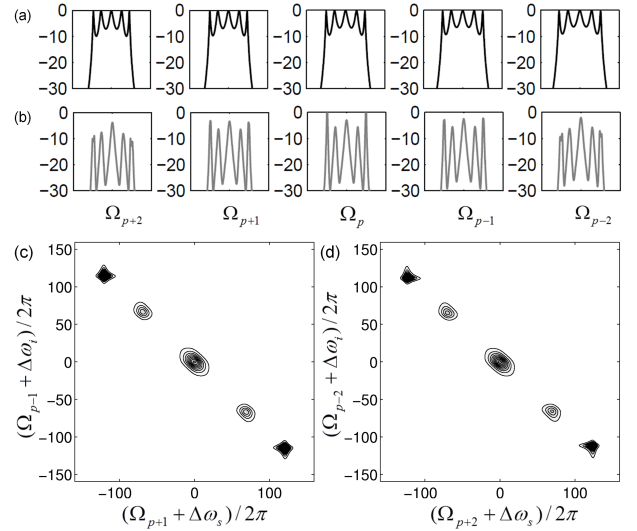}

\caption{(a) Spectra of the transmission bands of a coupled resonator waveguide consisting of five microrings. (b) Spectrum of the two photon state when a cw pump is placed at the resonance $\Omega_{p}$. (vertical axes are in logarithmic scale for both (a) and (b)) (c),(d) JSI of the transmission bands adjacent to the pump as well as two bands away. \label{fig:4}}
\end{figure}

While we have focused on the details of a single resonance in the prior discussion, as was predicted for for the case of a single resonator \cite{Chen2011}, the full two-photon state generated by the coupled resonator device is expected to form a "comb" structure with peaks centered around the resonance frequencies. In Fig. \ref{fig:4}(a) we plot the transmission spectra around five particular resonances of a 5-ring unapodized coupled resonator waveguide, taking into account both the dispersion of the intrinsic constituent waveguides as well as the dispersion of the directional couplers \cite{Aguinaldo2012}. The spectrum of the two photon state for a cw pump placed at the resonance $\Omega_{p}$ is given in Fig. \ref{fig:4}(b), showing a fine structure characteristic of the number of resonators. While the general structure remains consistent, the peaks near the edges are reduced more quickly than those near the middle. This can be attributed to the large directional coupler dispersion which give rise to non-uniform transmission bandwidths. Careful inspection of Fig. \ref{fig:4}(a) shows that the bandwidths increase gradually with frequency. The further apart the bands are, the more misaligned the transmission peaks become which in turn reduces the effective nonlinearity (see Eq. \ref{eq:6}), since transmission peaks correspond also to peaks in slowing factor. The band edge peaks are most adversely affected since they are also the narrowest. In Fig. \ref{fig:4}(c,d), we plot the JSI with signal and idler in the adjacent resonances as well as being two resonances apart from the pump. As compared to Fig. \ref{fig:3}(a), we can see that the band edge peaks have become more distorted. Clearly, the uniformity of the two photon state generated over the "comb" for the coupled resonator configuration is limited by the dispersion of the directional couplers, the suppression of which is a problem of interest not only for chip-scale quantum optics but in "classical" photonics as well. 

In summary, we have calculated the expected photon pair flux rates from a silicon
coupled-microring waveguide device based on spontaneous four-wave mixing, a nonlinear process which scales quadratically with the optical pump power. This hybrid structure may significantly outperform conventional waveguides of much longer length at realistic
waveguide losses and inter-resonator coupling strengths, and also outperform single ring resonators. We also developed
a quantum mechanical coupled-mode theory which may be applicable to a generic class of waveguide or resonator based integrated photonic quantum light source, and  evaluated the expected joint spectral intensities for the apodized
and unpodized cases. Spectral filtering to isolate individual Bloch
eigenmodes will help for heralding to a pure state. We also introduced a concept of 
tunability of the output Schimdt number, given control over individual
inter-resonator coupling coefficients. The special cases of flat transmission
and flat group delay may give nearly pure heralded states without need
for additional filtering. 

\begin{acknowledgments}
This work was supported by the National Science Foundation under grants
ECCS-0642603, ECCS-0925399, ECCS-1201308, NSF-GOALI collaboration with IBM, NSF-NIST
supplement, and UCSD-Calit2. J. R. Ong acknowledges support from Agency
for Science, Technology and Research (A{*}STAR) Singapore. 
\end{acknowledgments}
\bibliography{Quantum}
 
\end{document}